\documentclass[leqno,authoryear,1p]{article}
\usepackage[utf8]{inputenc}
\usepackage{xcolor}
\usepackage{times}
\usepackage{enumitem}
\usepackage{bbold}
\usepackage{dsfont}
\usepackage{comment}
\usepackage{mathrsfs}
\usepackage{slashed}
\usepackage[french,german,english]{babel}
\usepackage{soul}
\usepackage{amssymb,amsmath}
\usepackage{natbib}
\title{Einstein and the old quantum theory}
\author{Alexander Afriat}
\begin{document}
\maketitle

\begin{abstract}
\noindent Objecting that Sommerfeld's quantum conditions refer to particular coordinates, Einstein proposes a canonically invariant rule. But even if the invariance is canonical, Einstein may have in mind a double \emph{configuration} space invariance: with respect to loop deformations, and to \emph{point} transformations---all on a torus where features of the Liouville-Arnol'd theorem already appear.
\end{abstract}

\begin{quote}\begin{quote}
\tableofcontents
\end{quote}\end{quote}
\thispagestyle{empty}
\clearpage
\setcounter{page}{1}

\section{Introduction}

Einstein's role in quantum theory is well known: one associates him with the foundational debate in the twenties and thirties, with the photoelectric effect, perhaps with the quantum theory of gases---or even statistical mechanics---in general; but less with the old quantum theory of Bohr and Sommerfeld (from 1913), or with analytical mechanics for that matter. ``Zum Quantensatz von Sommerfeld und Epstein'' (1917a), which I propose to consider, has an unusual place in the history of science, characterised by neglect, limited attention then unexpected, retarded recognition. \citet{Gutzwiller} and \citet{Graffi} mention the peculiar history of its citations: practically half a century of silence,\footnote{Punctuated by two important wave-mechanical citations: \citet{Broglie} pp.~63ff, and the footnote on p.~495 of \citet{Zweite}. The story of how the \emph{mécanique ondulatoire} emerged---see \citet{Lanczos} pp.~277ff---as if by mistake from Louis de Broglie's remarkable misunderstandings of \emph{Quantensatz} deserves to be told, but elsewhere; and Schrödinger suggests, in the footnote, that \emph{Quantensatz} was behind his \emph{Wellenmechanik} as well.} then rediscovery\footnote{See \citet{Keller}, \citet{Gutzwiller} pp.~215, 282, 315, \citet{Spivak} p.~640.} in the more mathematical context of analytical mechanics, dynamical systems.

Even if analytical mechanics lay outside of Einstein's main mechanical interests---statistical mechanics, relativistic mechanics, foundations of mechanics---\emph{Quantensatz} can be treated more as analytical mechanics than as quantum theory. Sommerfeld's quantum rule\footnote{\citet{Sommerfeld1915} p.~429, \citet{Sommerfeld} p.~9; see also \citet{Epstein1916} p.~173, \citet{Epstein} pp.~490, 501.} ((\ref{Sommerfeld}) below), referred to in the title, becomes little more than a point of departure, almost a pretext for the development or at least adumbration of a futuristic, topological, highly invariant analytical mechanics; that at any rate is what it amounts to, however Einstein himself saw it.

In a nutshell, \citet{Quantensatz, Ehrenfest} objects that Sommerfeld's rule refers to particular coordinates, in the sense that there's a condition for each of the $l$ coordinates $q_1,\dots,q_l$. Einstein proposes another rule ((\ref{EinsteinRule}) below), which is more invariant inasmuch as it integrates the \emph{entire} momentum one-form $\mathbf{p}$---not one component $p_i$ at a time---over the $l$ homotopy classes that characterise the topology of the torus he eventually introduces (\S\ref{tworules} below). Which raises the issue of why Einstein's rule should be any better than Sommerfeld's. \emph{Empirical} superiority is often important in physics. Not here: Einstein never even brings it up, being only concerned with formal invariance; and even for us, favoured as we are by hindsight, empirical adequacy can hardly be invoked to compare theories we now know to be empirically very inadequate.\footnote{There is no reason why the two rules should agree on motions, ruling out the same ones. But even if such disagreement would make the rules significantly inequivalent, it would only give the preference to one if an independent (empirical) way of checking the selections were available; and we now know that an \emph{undulatory} quantum theory---with superposition, coherence, interference, resonance \emph{etc}.---is needed to make any sense at all of atomic mechanics, and select the right motions.} But perhaps there are cases where Einstein's rule makes more sense than Sommerfeld's, or works better. For an elliptical Kepler motion, Einstein's rule is no better than Sommerfeld's, both work well: the score there is \emph{one all}. And neither rule can really handle the self-intersections of a hypotrochoid:\footnote{Schrödinger writes \emph{Rosettenbahn}; in English one also sees \emph{rosetta orbit}. Einstein himself seems to have no term for such a self-intersecting orbit. We'll see that the exact shape of the figure is less important than its `topological' features: whether it intersects itself (unlike a mere ellipse), and whether it eventually closes.} 0-0 (cumulatively 1-1). The superiority of Einstein's rule will only emerge on the toroidal configuration space he constructs to resolve the self-intersections: 1-0 there (cumulatively 2-1). But even on the torus, the advantage is only qualified: all one can say is that Einstein's rule makes more sense and is more invariant, not that Sommerfeld's rule would make no sense at all.\footnote{Even if it would make no \emph{invariant} sense, no \emph{geometrical} sense, some unfortunate \emph{coordinate-dependent} sense could perhaps be salvaged.}

A notable, perhaps surprising, feature of Einstein's mechanics is its `Riemannian'---rather than symplectic---character: like Hamilton's own mechanics,\footnote{The distinction being between Hamilton's mechanics and \emph{Hamiltonian mechanics}, which is a symplectic mechanics in phase space, not configuration space.} it is firmly rooted in $l$-dimensional \emph{configuration} space\footnote{\citet{Quantensatz} p.~82: ``Da diese Formulierung von der Wahl der Koordinaten [$q_i$] nicht unabhängig ist, so kann sie nur bei bestimmter Wahl der Koordinaten zutreffen.'' See also p.~83, bottom; p.~84: ``in dem $l$-dimensionalen Raume der $q_i$ betrachten. Ziehe ich im Raume der $q_i$ [\,\dots].'' P.~85: ``der in Betracht kommende Raum der $q_i$''; ``im $q$-Raume''; ``ein Punkt $P$ des Koordinatenraumes mit den Koordinaten $Q_i$ [\,\dots].'' P.~86: ``jeden Punkt ($q_i$) des Koordinatenraumes''; ``im Koordinatenraume der $q_i$ [\,\dots].'' P.~87: ``im $q_i$-Raume''; ``in den $q_i$-Raum''; ``des $q_i$-Raumes''; and so on, page after page. There is a similar emphasis on configuration space (as opposed to phase space) in \citet{Ehrenfest,EinsteinJacobi}.\label{ConfigurationFootnote}} $\mathcal{Q}$, not in the $2l$-dimensional phase space $\mathit{\Gamma}=\mathbb{T}^*\mathcal{Q}$ we're used to today. Geometric mechanics nowadays is typically symplectic, but not always---books by \citet{CalinChang} and \citet{Pettini} are exceptions. Today's symplectic abstractions would have been foreign, even unintelligible, to almost everyone\footnote{Except a handful of mathematicians (Whittaker, Levi-Civita, Carathéodory \emph{etc.}), and perhaps the most mathematical physicists (Born, Schrödinger).} in 1917; and Riemannian geometry is much closer to the curved Lorentzian geometry Einstein had practically invented and was still immersed in at the time. The presence of the Hamiltonian $\mathscr{H}$ and other functions of position \emph{and indeed momentum} in Einstein's formalism is misleading; even such objects have to be situated in configuration space: he thinks of the Hamiltonian as assigning \emph{a function}
$$
\mathscr{H}_{\mathbf{q}}\textrm{(}\mathbf{p}\textrm{)}:\mathbb{T}^*_\mathbf{q}\mathcal{Q}\rightarrow \mathbb{R}
$$
\emph{of momentum} $\mathbf{p}$ to every $\mathbf{q}$ in configuration space $\mathcal{Q}$---rather than a real number $\mathscr{H}\textrm{(}\mathbf{z}\textrm{)}$ to every point $\mathbf{z}=\textrm{(}\mathbf{q},\mathbf{p}\textrm{)}\in\mathit{\Gamma}$ of phase space. Today we understand integrability in phase space; Einstein tries to understand it in the configuration space in which one even has to situate his gropings (\S\ref{Liouville} below) towards the Liouville-Arnol'd theorem.

A fixed configuration space $\mathcal{Q}$ makes Einstein's mechanics more Lagrangian than Hamiltonian---even if he assigns momenta $\mathbf{p}\in \mathbb{T}^*_\mathbf{q}\mathcal{Q}$ rather than velocities $\dot{\mathbf{q}}\in \mathbb{T}_\mathbf{q}\mathcal{Q}$ to the various points $\mathbf{q}$ of $\mathcal{Q}$. The whole point of Hamiltonian mechanics being to go beyond mere point transformations, which refer to a fixed configuration space $\mathcal{Q}$; its added generality is contained in the exact term $dF$ in (\ref{symplectic}) and (\ref{exactterm}) below.

Einstein seems to have in mind a double \emph{configuration} space invariance ((\ref{twofreedoms}) below): with respect to loop deformations and \emph{point} (as opposed to \emph{canonical}) transformations. \citet{Graffi} extends the latter invariance from configuration space to phase space. The integral (\ref{integral}) below \emph{is} canonically invariant, that's undeniable; the issue is Einstein's \emph{understanding} of the invariance---whether he (or any physicist in 1917) thought in genuinely symplectic terms. Symplectic geometry as we know it has been around for no more than a few decades; Einstein in any case was no symplectic geometer.\color{black}

``Integrability'' can mean various things. There is the purely geometrical notion, well exemplified by the exactness of a one-form $\alpha=df$: infinitesimal objects $\alpha\textrm{(}\mathbf{q}\textrm{)}$ assigned to every point $\mathbf{q}$ fit together in such a way as to allow `derivation from a potential $f$,' as Einstein would put it. But that's abstract; other notions are more concretely arithmetical, and have to do with the simplicity or even the very possibility of coordinate representation. The numbers involved\footnote{In \S\ref{Liouville} below there will be the conserved quantities $f=(f_1,\dots,f_l)\in\mathbb{R}^l$.} acquire geometrical meaning by identifying submanifolds, which can simplify the representation of (integrated) motion by \emph{adaptation}---by `following it so as to eliminate it,' absorbing its twists \& turns into their adapted shapes, thus enclosing and therefore representing it.

Einstein has his own notion of integrability, which is so strong it goes well beyond what is now known as `complete' integrability: it involves confining the motion to a one-dimensional (`closed') manifold, of finite length, a loop---``die Bahn ist dann eine geschlossene, ihre Punkte bilden ein Kontinuum von nur einer Dimension.''\footnote{\citet{Quantensatz} p.~88} He seems to reduce integrability and separability\footnote{See \citet{Gutzwiller} \S3.7 on the \emph{modern} distinction (not Einstein's).} to two simple `topological' distinctions: whether the motion eventually closes, or never does (which would pose an \emph{irremediable} problem for quantisation); and if it does close, whether it intersects\footnote{Intersections can only occur in configuration space ($q_i$-\emph{Raum}), not phase space; see footnote \ref{ConfigurationFootnote}.} itself (which poses a \emph{remediable}, indeed most welcome, problem for quantisation---so welcome that Einstein's whole invariant, topological agenda rests on it). Again, self-intersections are remedied by \emph{Riemannisierung}, in other words enlargement of the configuration space, considered in the next section.

\section{The two quantum rules}\label{tworules}
Sommerfeld's quantum conditions
\begin{equation}\label{Sommerfeld}
J_i^*=\oint p_idq_i=n_ih\textrm{,}
\end{equation}
$i=1,\dots,l$, rule out atomic motions whose actions $J_i^*$ are not integer multiples $n_i\in\mathbb{Z}$ of Planck's constant $h$. Having spent the previous years immersed in the tensorial covariance of general relativity, Einstein sees a problem here: Sommerfeld's rule refers to the specific coordinates $q_1,\dots,q_l$; each one of the $l$ conditions concerns a particular momentum component $p_i$. Einstein replaces them with $l$ conditions, each one of which involves the \emph{entire} momentum one-form
\begin{equation}\label{momentum}
\mathbf{p}=\sum_{i=1}^lp_idq_i\textrm{,}
\end{equation}
now integrated over the $l$ homotopy classes $\mathbb{H}_1,\dots,\mathbb{H}_l$ or `topological features' of the space---there being a quantum condition for every such feature. To understand the construction we can begin with an annulus.

Einstein considers the example of an annulus $\mathfrak{Q}$ bounded by circles of radii $r_1$, $r_2$. Motions on $\mathfrak{Q}$ that never close are \emph{in}tractable; those that close without intersecting themselves are \emph{too} tractable---for Sommerfeld's rule in particular, thereby giving Einstein no advantage. Einstein's whole strategy relies on closed motions that intersect themselves: the only motions combining the two virtues of being somehow tractable, but not so much as to preclude the welcome difficulties in whose solution lies Einstein's real edge over Sommerfeld, and which represent the chief interest of \emph{Quantensatz}.

To see the problem, the above `topological' classification of motions (periodic or not, self-intersecting or not) can be re-expressed in terms of momentum assignments to points, or rather \emph{Stellen}\footnote{To include space-filling (non-periodic) motion in the classification Einstein broadens his purview from the point to its immediate surroundings.} (which are somewhat larger): [1] \emph{one}, [2] \emph{finitely many}, [3] \emph{infinitely many}. Einstein considers an infinitesimal region$\mathfrak{r}\subset \mathfrak{Q}$ crossed by the motion, the following cases can arise:
\begin{enumerate}[label={[\arabic*]}]
\item The next time the motion crosses $\mathfrak{r}$ it assigns to $\mathbf{q}\in\mathfrak{r}$ the same momentum $\mathbf{p}$---and hence every time thereafter. This is the simplest kind of periodicity: the motion is closed (and hence confined to a loop, a one-dimensional manifold of finite length) and never intersects itself.
\item The motion eventually assigns a \emph{finite} number $N$ of momenta\footnote{Each $\mathbf{p}_k\in\bigwedge^1\mathcal{Q}$ is a one-form on $\mathcal{Q}$ and not a coordinate $\in\mathbb{R}$.} $\mathbf{p}_1,\dots,\mathbf{p}_N$ to $\mathfrak{r}$, finally closing on the $N$-th lap. The \emph{Bahn} is still \emph{eine geschlossene, ihre Punkte bilden ein Kontinuum von nur einer Dimension}; but it intersects itself.\footnote{See \citet{Gutzwiller} figures 10 \& 30.}
\item The motion assigns an \emph{infinite} number of momenta to $\mathfrak{r}$, without ever closing. As the motion is not periodic, it cannot be confined to a loop, an \emph{exakt geschlossene Bahn}.
\end{enumerate}

The quantisation rules at issue here only make sense with a single momentum at a \emph{Stelle}. To make the momentum assignment amenable to quantisation, Einstein enlarges the configuration space, thus restoring single-valuedness. But the enlargement procedure he adopts is \emph{finite}, and cannot be repeated \emph{infinitely} many times (for infinitely many momentum values). \emph{Riemannisierung} requires periodic motions.

So Einstein takes a closed hypotrochoid, which intersects itself. He chooses a point $\mathbf{q}\in\mathfrak{Q}$ to which the motion assigns two momenta, $\mathbf{p}_1$ and $\mathbf{p}_2$. In order to restore the \emph{Einwertigkeit} needed for quantisation he superposes a second annulus on the first, identifying the delimiting circles, and stipulating that whenever the motion reaches either one it changes annulus. The motion on the resulting two-torus $\mathfrak{T}^2$ can now be quantised since it no longer intersects itself. The topology of the torus is captured by the (nontrivial) homotopy classes\footnote{\citet{Graffi} p.~23: ``curve topologicalmente inequivalenti.''} $\mathbb{H}_1$ and $\mathbb{H}_2$, respectively made up of loops going around the first and second circles of the torus (once).\footnote{See \citet{Gutzwiller} figure 30.} The integral
\begin{equation}\label{integral}
\langle\mathbf{p},\mathbb{H}_i\rangle=\oint_{\mathbb{H}_i}\hspace{-5pt} \mathbf{p}
\end{equation}
vanishes for neither $\mathbb{H}_1$ nor $\mathbb{H}_2$, whereas $\langle\mathbf{p},\mathbb{H}_0\rangle$ does vanish for the trivial homotopy class $\mathbb{H}_0$ of `contractible' loops going around neither circle: Einstein specifies that the one-form $\mathbf{p}$ on $\mathfrak{T}^2$ is $\kappa$losed\footnote{Since I am using ``closed'' in two entirely different senses, I'll write ``$\kappa$losed'' for this second sense (`locally exact'), pertaining to differential forms.} (or perhaps even exact---see \S\ref{HamiltonJacobi} below). The new quantum rule\footnote{See \citet{Graffi04} p.~175.}
\begin{equation}\label{EinsteinRule}
\langle\mathbf{p},\mathbb{H}_i\rangle=n_i h
\end{equation}
$i=1,\dots,l$, which in fact applies more generally to any $l$-torus\footnote{In the paragraph on p.~90 (1917a) containing figures 1 \& 2 Einstein seems to have in mind an $l$-dimensional \emph{torus} $\mathfrak{T}^{l}$ (rather than a more general manifold with Betti number $l$). In two dimensions, \emph{Riemannisierung} clearly produces a torus; little generality is lost in considering $\mathfrak{T}^2$; and one wonders how the scheme can work in general if the enlarged configuration space is not toroidal. So I'll speak of a torus $\mathfrak{T}^{l}$ even with $l>2$.} $\mathfrak{T}^{l}$, requires the loop integrals to be integer multiples $n_i\in\mathbb{Z}$ of Planck's constant $h$. The methodological, æsthetic superiority of the rule lies in the double invariance of the expression
$$
\oint_{\mathbb{H}_k}\sum_ip_idq_i:
$$
invariance with respect to the choice of coordinates $q_i$, and also with respect to the particular loop of $\mathbb{H}_k$---with respect to `loop deformations.'\footnote{\emph{Cf}.\ \citet{Graffi} p.~22.} All `old quantum theories' are empirically too inadequate to warrant an \emph{empirical} preference of Einstein's rule over Sommerfeld's.

Again, both quantum rules work in case [1], neither one in cases [2] and [3]; but at least in case [2], the multi-valuedness (being finite) can be resolved on a larger configuration space $\mathfrak{T}^l$, where Einstein's rule works well and makes perfect sense. Even Sommerfeld's conditions can be \emph{made to} work on $\mathfrak{T}^l$---which, however, strongly favours the double invariance of Einstein's rule.

\section{Loops, \emph{trajectoires}, gauge}
Einstein's loops are meaningless on their own and only acquire significance collectively as elements of the homotopy classes $\mathbb{H}_1,\dots,\mathbb{H}_l$ which capture the topological peculiarities of the enlarged configuration space. Even if Einstein goes to the trouble of writing ``\foreignlanguage{german}{irgendeine geschlossene Kurve, welche durchaus keine "`Bahnkurve"' des mechanischen Systems zu sein braucht},'' Broglie nonetheless speaks of ``\emph{trajectoires} fermées'' in the chapter of his \emph{Thèse} (1924, p.~23) he devotes to \emph{Quantensatz}. But Broglie's misunderstanding is in fact more interesting than one may imagine, enough to deserve a few words.

Consider such a mechanical \emph{trajectoire}\footnote{If the Hamiltonian has no explicit dependence on time, the \emph{trajectoire} is best viewed as a mere (one-dimensional) manifold, since the progression of its (temporal) parameter would be trivial; see \S\ref{HamiltonJacobi} below.} $\xi$ in the plane $\mathbb{P}$, and the action integral
\begin{equation}\label{Cartan}
J^*=\oint_{\xi}(p_1dq_1 +p_2dq_2)
\end{equation}
calculated with respect to the coordinates $q_1,q_2$. To simplify we can confine the `source'\footnote{The term ``source'' comes from the `divergence' version of Stokes's theorem (attributed to Gauß or Ostrogradsky or perhaps Green), here it is more metaphorical. Here the `source' at the origin produces a turbulence which by the theorem manifests itself on the loop $\xi$ as the circulation $J^*$.} to the origin $\mathbf{0}\in\mathbb{P}$; in other words the curl $d\mathbf{p}$ of the momentum one-form
$$
\mathbf{p}=p_1dq_1+p_2dq_2
$$
vanishes everywhere else, on all of $\bar{\mathbb{P}}=\mathbb{P}-\{\mathbf{0}\}.$ The integral (\ref{Cartan}) has various interesting symmetries\footnote{See \citet{Epstein1916} p.~172, \citet{Whittaker} pp.~271ff, \citet{LeviCivita} pp.~353ff.} which are worth looking at. A diffeomorphism
$$
\gamma: \mathbb{P}\rightarrow\mathbb{P}\hspace{2pt};\hspace{4pt}\mathbf{q}\mapsto \mathbf{Q}=\gamma\textrm{(}\mathbf{q}\textrm{)}\hspace{2pt};\hspace{4pt}\xi\mapsto\Xi=\gamma\textrm{(}\xi\textrm{)}
$$
defined on all of $\mathbb{P}$ displaces everything---\emph{trajectoire}, source at the origin, and other points $\bar{\mathbb{P}}$---so as to preserve the relations of inclusion and exclusion, and hence the integral $J^*$ itself. Even a diffeomorphism $\bar{\gamma} : \bar{\mathbb{P}}\rightarrow \bar{\mathbb{P}}$ on $\bar{\mathbb{P}}$ (as opposed to $\mathbb{P}$)---which only displaces points where $d\mathbf{p}$ vanishes, and the \emph{trajectoire} itself, but not the source $\mathbf{0}$---would be just as symmetric, as it could never drag $\xi$ over the source. For a diffeomorphism to alter the topological relations on which (\ref{Cartan}) depends it would have to displace \emph{selectively}, telling apart origin and points of $\xi$.

Since $\gamma$ pulls a real-valued function
$$
Q_i:\mathbb{P}\rightarrow\mathbb{R}\hspace{2pt};\hspace{4pt}\mathbf{Q}\mapsto Q_i\textrm{(}\mathbf{Q}\textrm{)}
$$
defined on the range $\mathbb{P}$ back\footnote{The case is so trivial (domain and range coincide \emph{etc}.) that pulling back and pushing forward can be legitimately confused; so the abuse of notation $\gamma^{\circledast}\mathbf{p}$, for instance, is venial. A diffeomorphism can be taken to map from the domain $\mathbb{P}$ to the range $\mathbb{P}$, or the other way around.} to the domain $\mathbb{P}$ of $\gamma$, yielding a function
$$
q_i=\gamma^{\circledast} Q_i=Q_i\circ \gamma:\mathbb{P}\rightarrow\mathbb{R}\hspace{2pt};\hspace{4pt}\mathbf{q}\mapsto q_i\textrm{(}\mathbf{q}\textrm{)}=(\gamma^{\circledast} Q_i)\textrm{(}\mathbf{q}\textrm{)}
$$
on the domain $\mathbb{P}$, it pulls the differential $dQ_i$ back accordingly---and hence the basis $dQ_1,dQ_2$, and with it the linear combination:
$$
\mathbf{p}=\gamma^{\circledast}\mathbf{P}=p_1\gamma^{\circledast}dQ_1+p_2\gamma^{\circledast}dQ_2\textrm{,}
$$
where I have written $\circledast$ to avoid confusion with the asterisk Einstein uses to denote the Maupertuis action $J^*$. We can write
$$
J^*=\oint_{\xi}\mathbf{p}=\oint_{\Xi}\mathbf{P}\textrm{,}
$$
or even
$$
J^*=\oint_{\Xi}\mathbf{p}=\oint_{\xi}\mathbf{P}\textrm{,}
$$
and
\begin{align}\label{twofreedoms}
\begin{split}
J^*&=\oint_{\gamma'\textrm{(}\xi\textrm{)}}\hspace{-2pt}\gamma^{\circledast}\mathbf{p}\\
&=\oint_{\mathbb{H}}\gamma^{\circledast}\mathbf{p}=\langle\mathbf{\gamma^{\circledast}p} ,\mathbb{H}\rangle
\end{split}
\end{align}
for all diffeomorphisms
$$
\gamma,\gamma': \mathbb{P}\rightarrow\mathbb{P}
$$
leaving the source inside the \emph{trajectoire}. Broglie's \emph{trajectoires} thus become highly transformable, invariant entities---little more than \emph{deformable loops expressive of topological properties}. It remains a mistake to think of Einstein's loops $\xi\in\mathbb{H}$ as \emph{trajectoires}, but not an uninteresting one.

In Hamiltonian mechanics one distinguishes between \emph{point} transformations $\gamma$ (on the $l$-dimensional $q_i$-\emph{Raum}) and \emph{canonical} transformations (on the $2l$-dimensional phase space $\mathit{\Gamma}$).\footnote{See \citet{Landau} \S~45.} So far we haven't gone beyond the rather restrictive `point' condition
\begin{equation}\label{pointfreedom}
\mathbf{p}=\sum_ip_idq_i=\sum_iP_idQ_i
\end{equation}
of Lagrangian mechanics (where the configuration space $\mathcal{Q}$ characterised by $q_1,\dots,q_l$ \emph{maintains its identity}, without `getting lost' in the $2l$-dimensional state space); but in fact the weaker `symplectic' condition
\begin{align}\label{symplectic}
\begin{split}
\omega&=d\mathbf{p}=d\mathbf{p}'=d(\mathbf{p}+dF)\\
&=\sum_idp_i\wedge dq_i=\sum_idp_i\wedge dq_i+d^2F
\end{split}
\end{align}
also preserves $J^*$, where
\begin{equation}\label{exactterm}
\mathbf{p}\mapsto \mathbf{p}'=\mathbf{p}+dF
\end{equation}
and the generating function $F$ is a zero-form. So in addition to the diffeomorphic freedom $\gamma$ (or $\bar{\gamma}$) and the homotopic freedom $\xi\mapsto \Xi$ (keeping clear of the source), we have the `gauge' or `symplectic' freedom (\ref{symplectic}) to add an exact term $dF$:
$$
J^*=\langle\gamma^{\circledast}\mathbf{p} +dF,\mathbb{H}\rangle.
$$
While a diffeomorphism $\gamma$ affects the momentum one-form indirectly by first dragging points, the gauge transformation (\ref{exactterm}) is fibre-preserving and therefore acts directly on each $\mathbf{p}\textrm{(}\mathbf{q}\textrm{)}$, point by point. The canonical transformation generated by $F$ also corresponds to a deformation of curves in $\bar{\mathbb{P}}$, reminiscent of the loop deformation $\xi\mapsto \Xi$. Even if the curl at the origin (or the corresponding circulation $J^*$) prevents the one-form $\mathbf{p}$ from having a \emph{global} primitive, one can think of a \emph{local} primitive $\lambda$ satisfying $\mathbf{p}=d\lambda$ locally.\footnote{``Locally'' could mean, for instance, \emph{on any simply-connected region of $\bar{\mathbb{P}}$}.} In much the same way as one can take $\xi$ to be displaced by the point transformation $\gamma$, one can take the level sets $\lambda=\textrm{const}.$ to be deformed by $F$.

\citet{Graffi} has rightly pointed out that Einstein's rule (\ref{EinsteinRule}) is \emph{canonically} invariant. The symplectic freedom (\ref{symplectic}) is undeniably \emph{available}; but Einstein's mechanics is so \emph{un}symplectic, so firmly rooted in the $q_i$-\emph{Raum} he so often mentions, that I doubt he had in mind anything beyond the two genuinely $q_i$-\emph{Raum} symmetries represented in (\ref{twofreedoms}).

\section{Hamilton-Jacobi theory}\label{HamiltonJacobi}

The old quantum theory, including \emph{Quantensatz}, was formulated in terms of Hamilton-Jacobi theory,\footnote{See \citet{Hamilton1833, Hamilton1834}, \citet{Jacobi84} pp.~143ff, \citet{Appell1} pp.~556ff, \citet{Epstein} pp.~493ff, \citet{Whittaker} pp.~288ff, \citet{Caratheodory} pp.~66ff, \citet{Brillouin} pp.~168ff, \citet{Appell} pp.~429ff, \citet{LeviCivita} pp.~355ff, \citet{Arnold} \S9.4, \citet{Jacobi96} pp.~216ff, \citet{FasanoMarmi} \S11.1, \citet{Graffi04} p.~51, \citet{BenciFortunato} \S1.4.} which therefore deserves some attention.

The principle of least action\footnote{See \citet{Brillouin} pp.~159ff.} exists in two versions, which in Einstein's notation would be distinguished by a star indicating neglect of time.

\begin{enumerate}[label={[\roman*]}]
\item The `space-time' Hamiltonian version determines the spatial shape of trajectories \emph{as well as motion along them} by minimising the full Hamiltonian action\footnote{\emph{Cf}.\ \citet{Caratheodory} p.~10.}
$$
J=\int_{t_0}^{t}\hspace{-2pt} \mathscr{L}\,dt\textrm{,}
$$
where the `momentum' or `covariant' Lagrangian can be written
$$
\mathscr{L}=\sum_i p_i\frac{\partial \mathscr{H}}{\partial p_i}-\mathscr{H}.
$$
\item The `spatial' version attributed to Maupertuis only gives the spatial shape of the trajectory, by minimising the purely spatial part
\begin{equation}\label{Maupertuis}
J^*=\int_{\mathbf{q}_0}^{\mathbf{q}}\hspace{-2pt}\mathbf{p}=\int_{\mathbf{q}_0}^\mathbf{q}\sum_i p_idq_i
\end{equation}
of the action.
\end{enumerate}
Hamilton's principle is in a sense more general; but the generality it adds to Maupertuis' is only of any interest if the Hamiltonian $\mathscr{H}$ depends explicitly on time. Since Einstein takes it not to, the motion is confined to a level surface where the Hamiltonian remains equal to some constant $E$; and the action assumes the degenerate additive form\footnote{See \citet{Jacobi84} \emph{Einundzwanzigste Vorlesung}, \citet{Appell} pp.~430ff, \citet{LeviCivita} pp.~362ff.}
$$
J=J^*-Et.
$$
Let us write $\mathscr{H}=T+U$ (and $\mathscr{L}=T-U$), where the potential $U$ depends only on position and the kinetic energy
\begin{align*}
T=\frac{\|\mathbf{p}\|^2}{2m}&=\frac{1}{2}\sum_i p_i\dot{q}_i\\
&=\frac{1}{2}\sum_i p_i\frac{\partial \mathscr{H}}{\partial p_i}
\end{align*}
is quadratic in the momenta. Only the spatial shape of the trajectory would then remain interesting. To confine our attention to that shape, ignoring the trivial time evolution given by the term $-Et$, we just take the spatial `Maupertuis' part of $J$, namely (\ref{Maupertuis}). Viewing $\mathbf{q}_0$ as a fixed initial position and $\mathbf{q}$ as a variable final position, we can radiate $\mathbf{q}$ in all directions from $\mathbf{q}_0$ (now reminiscent of a source in geometrical optics\footnote{See \citet{Hamilton1833, Hamilton1834}, \citet{Whittaker} pp.~288ff.}) along dynamical trajectories derived from the action function $J^*\textrm{(}\mathbf{q},\mathbf{q}_0\textrm{)}$ satisfying the Hamilton-Jacobi equation\footnote{See \citet{Graffi04} p.~51.}
\begin{equation}\label{HJequation}
\mathscr{H}(\mathbf{q},dJ^*)=E\textrm{,}
\end{equation}
or rather
$$
\|\mathbf{p}\|^2=\|dJ^*\|^2\\=2m(E-U).
$$
Choosing an infinitesimal action $\delta J^*$, we first have the infinitesimal sphere $\sigma(\mathbf{q}_0,\delta J^*)$ of radius
$$
\zeta(\mathbf{q}_0,\delta J^*)=\frac{\delta J^*}{\sqrt{2m(E-U\textrm{(}\mathbf{q}_0\textrm{)})}}
$$
around $\mathbf{q}_0$. The momentum $\mathbf{p}$ at any $\mathbf{q}$ on the sphere is the differential
$$
\mathbf{p}\textrm{(}\mathbf{q}\textrm{)}=dJ^*\textrm{(}\mathbf{q}\textrm{)}=\sum_{i=1}^l\frac{\partial J^*\textrm{(}\mathbf{q}\textrm{)}}{\partial q_i}dq_i
$$
of $J^*$ viewed as a function of the final position $\mathbf{q}$. Once we have a level surface $\sigma(\mathbf{q}_0,\delta J^*)$ of $J^*$, we might as well generalise and propagate from an \emph{arbitrary} ($l-1$)-dimensional initial surface
$$
\Sigma_1=\Sigma_{J_{[1]}^*}
$$
(which could be a sphere $\sigma$ or not), viewed as a level surface of action $J^*_{[1]}$. I'll follow \citet[pp.~492-3]{Zweite} on the way to his \emph{Wellenmechanik}.\footnote{See also \citet{Brillouin} pp.~169ff, \citet{Arnold} pp.~251ff.} We can again take the same increment\footnote{A smaller increment would be even better, if secondary waves are propagated from an \emph{already infinitesimal} primary wavefront $\sigma(\mathbf{q}_0,\delta J^*)$.} $\delta J^*$, which provides a distance $\zeta (\mathbf{q}_1,\delta J^*)$ at the generic point $\mathbf{q}_1\in\Sigma_1$. Then there are two possible constructions: Either we repeat the above construction, now treating $\mathbf{q}_1$ as the source of a `secondary' wave,\footnote{See \citet{Huygens}, \citet{Whittaker} pp.~289ff, \citet{Caratheodory} pp.~13ff, \citet{Arnold} \S9.3.} a sphere $\sigma(\mathbf{q}_1,\delta J^*)$ of radius $\zeta (\mathbf{q}_1,\delta J^*)$ emanated from every point of the `primary' wavefront $\Sigma_1$, in which case the surface $\Sigma_2$ of action
$$
J_{[2]}^*=J_{[1]}^*+\delta J^*
$$
is the envelope of the secondary waves. Alternatively one lays off $\zeta (\mathbf{q}_1,\delta J^*)$ normally at every point, to define $\Sigma_2$ even more directly. `Normally' can be understood as follows: The ($l-1$)-dimensional linear space
$$
\mathbb{T}_{\mathbf{q}_1}\Sigma_1\subset \mathbb{T}_{\mathbf{q}_1}\mathcal{Q}
$$
tangent to $\Sigma_1$ at $\mathbf{q}_1$ determines a ray $\rho_1^{\flat}$ in the cotangent space $\mathbb{T}^*_{\mathbf{q}_1}\mathcal{Q}$. The direction of the momentum $\mathbf{p}_1$ at $\mathbf{q}_1$ is given by $\mathbf{p}_1\in \rho_1^{\flat}$, the length by
$$
\|\mathbf{p}_1\|=\sqrt{2m(E-U\textrm{(}\mathbf{q}_1\textrm{)})}.
$$
The inverse $m^{\sharp}$ of the mass tensor\footnote{See \citet{Brillouin} p.~143.} $m^{\flat}$ determines a vector $\mathbf{p}_1^{\sharp}=m^{\sharp}\mathbf{p}_1$, indeed a ray $\rho_1^{\sharp}$ containing $\mathbf{p}_1^{\sharp}$ in the tangent space $\mathbb{T}_{\mathbf{q}_1}\mathcal{Q}$. The action increment $\delta J^*$ fixes the length $\zeta (\mathbf{q}_1,\delta J^*)$ of the required vector $\hat{\mathbf{p}}_1^{\sharp}\in\rho_1^{\sharp}$, which goes from $\Sigma_1$ to the corresponding point of $\Sigma_2$.

So that's one way (specificially a rather \emph{Hamiltonian} way) of understanding the Hamilton-Jacobi equation (\ref{HJequation})---which is so central to the old quantum theory, and even to wave mechanics: as an infinitesimal condition governing the orthogonal propagation in configuration space $\mathcal{Q}$ of a solution $J^*$ from an initial surface (which could be as small as a point). But the orthogonal construction also serves to elucidate, by opposition, the somewhat different logic of the `oblique' construction Einstein proposes in \S~3 of \emph{Quantensatz}. In Schrödinger's construction, the direction of propagation from the initial surface $\Sigma_1$ is determined by the local slant of the surface itself,\footnote{Here the normal direction of propagation is determined by the mass tensor $m$. With a more general Hamiltonian (and Lagrangian), the duality relations $\sharp\,\,\flat$ between level surface and motions are given by the fibre derivatives, with components ${\partial \mathscr{H}}/{\partial p_i}$, ${\partial \mathscr{L}}/{\partial\dot{q}_i}$.} not by initial data freely assigned to it, point by point.
Einstein also propagates from an ($l-1$)-dimensional surface in the $q_i$-\emph{Raum} $\mathcal{Q}$, but \emph{obliquely}---at an `angle,' relative to the surface, that varies from point to point. Schrödinger's orthogonal propagation, which makes the initial surface $\Sigma_1$ a \emph{level} surface of $J^*$, also makes the propagated field \emph{exact}, indeed consistent with a Hamilton-Jacobi potential $J^*$; Einstein's `oblique' construction has to require exactness independently, as it is not provided by orthogonality.

The difference between $\kappa$\emph{losed} and \emph{exact}, it should be noted, may not have been entirely clear to Einstein (or others in 1917). Important progress is made in Rham's \emph{Thèse} (1931), where the distinction is used for \emph{analysis situs}. Even if the distinction is clear to \citet{Weyl}, he confusingly uses the word ``exact'' to mean $\kappa$\emph{losed}: ``A differential $\omega$ whose derivative vanishes is called \emph{exact}.'' Weyl's understanding of $\kappa$losed is already rather modern:
\begin{quote}
What we mean by $\omega\sim 0$ ($\omega$ homologous zero) may be explained in two ways: either differentially as indicating that $\omega$ is the derivative of a differential of next lower rank, or integrally as demanding that the integral of $\omega$ over any cycle vanishes. Every differential $\sim 0$ is exact; one readily proves this in both ways. ``In the small'' both notions, exact and $\sim 0$, coincide, but not in the large.
\end{quote}
But one can imagine how roughly, if at all, the distinction was understood in 1917. In (1917c) Einstein seems to treat his equations (7) and (7a) as equivalent; in \emph{Quantensatz} he writes ``bzw.'' between his equations 10) and 10a)---but then does draw a related distinction, between \emph{einwertig} and \emph{vielwertig}, on the next page:
\begin{quote}
Ist aber der in Betracht kommende Raum der $q_i$ ein mehrfach zusammenhängender, so gibt es geschlossene Bahnen, welche nicht durch stetige Änderung auf einen Punkt zusammengezogen werden können; ist dann $J^*$ keine einwertige (sondern eine $\infty$ vielwertige) Funktion der $q_i$, so wird das Integral [$\int \sum p_i dq_i$] für eine solche Kurve im allgemeinen von Null verschieden sein.
\end{quote}
Indeed if the one-form $\mathbf{p}$ is only
\begin{quote}
$\mathbf{CMC}$: \emph{closed on a multiply-connected region},
\end{quote}
it may or may not be exact, the loop integral (\ref{integral}) may or may not vanish. If $\mathbf{p}$ is not exact,\footnote{The `topological interpretation' of the Aharonov-Bohm effect shows how important it is to distinguish between \emph{not} exact  and the above condition $\mathbf{CMC}$: not \emph{necessarily} exact.} the loop integral $\langle \mathbf{p},\mathbb{H}\rangle$ will not vanish, but its \emph{Vielwertigkeit} will be the \emph{denkbar einfachste}: every point $\mathbf{q}$ on a given loop acquires an additional $\langle \mathbf{p},\mathbb{H}\rangle$ on every lap:
\begin{align*}
J_{\textrm{(}n+1\textrm{)}}^*\textrm{(}\mathbf{q}\textrm{)}&=J_{\textrm{(}n\textrm{)}}^*\textrm{(}\mathbf{q}\textrm{)}+\langle \mathbf{p},\mathbb{H}\rangle\\
&=J_{\textrm{(}1\textrm{)}}^*\textrm{(}\mathbf{q}\textrm{)}+n\langle \mathbf{p},\mathbb{H}\rangle\textrm{,}
\end{align*}
where the integer $n$ stands for the lap. In terms of the \emph{Ein/Vielwertigkeit} of the primitive $J^*$, \emph{exact} means \emph{has a single-valued global primitive $J^*$}; merely $\kappa$\emph{losed} means \emph{exact ``in the small''}---the primitive $J^*$ is \emph{locally einwertig} but may be globally \emph{vielwertig}.

Returning to Einstein's construction, he freely assigns momenta $\textbf{p}_1\textrm{(}\textbf{q}_1\textrm{)}$ to the points $\textbf{q}_1$ of $\Sigma_1$, then radiates dynamical (Hamiltonian) trajectories\footnote{The almost superfluous curves $L$ and vectors $\dot{L}$ used to extend the one-form $\mathbf{p}_0$ on $\Sigma_0$ to a one-form $\mathbf{p}$ on the $l$-dimensional region $\mathcal{U}$ are only introduced because Einstein seems to have such objects in mind. They can easily be dispensed with, the one-forms are enough.} $L$ satisfying
$$
\dot{L}\textrm{(}\textbf{q}_1\textrm{)}=\mathbf{p}_1^{\sharp}\textrm{(}\textbf{q}_1\textrm{)}=m^{\sharp}\mathbf{p}_1\textrm{(}\textbf{q}_1\textrm{)}
$$
throughout $\Sigma_1$. The dynamical vector field $\dot{L}$ thus determined by the one-form $\mathbf{p}_1=\mathbf{p}|_{\Sigma_1}$ provides a one-form $\mathbf{p}=m^{\flat}\dot{L}$ on an $l$-dimensional region $\mathcal{U}\subset \mathcal{Q}$ covered by $\dot{L}$.

Einstein then wonders when the one-form $\mathbf{p}$ determined by the Hamiltonian vector field $\dot{L}$ will also satisfy the Hamilton-Jacobi equation; for the radiated congruence can, confusingly, be `Hamiltonian' (in other words \emph{made up of dynamical trajectories}) without being `Hamilton-Jacobi'---consistent with a potential $J^*$ satisfying the Hamilton-Jacobi equation. It turns out that $\mathbf{p}_1$ has to be $\kappa$losed, for then $\mathbf{p}$ will be too; and as long as $\mathcal{U}$ is topologically simple, $\kappa$losed means exact:
$$
(d\mathbf{p}_1=0)\Leftrightarrow (d\mathbf{p}=0)\Leftrightarrow (\exists\,J^* :\mathbf{p}=dJ^*).
$$
The potential $J^*$ is slightly underdetermined by its derivative $dJ^*$, in other words
$$
d^{-1}\mathbf{p}=[J^*]=[J^*+\eta]_{\eta}\textrm{,}
$$
where $\eta$ is a constant on $\mathcal{U}$. A value $J^*\textrm{(}\mathbf{q}\textrm{)}$ at a single $\mathbf{q}\in \mathcal{U}$ is enough to overcome the underdetermination and fix all of $J^*$. Summing up, the theorem can be given as follows: Only a $\kappa$losed momentum field $\mathbf{p}$ (radiated \emph{dynamically} from the momenta $\mathbf{p}_1$ freely assigned to an initial surface $\Sigma_1\subset \mathcal{Q}$) can be derived from a potential
$$
J^*=d^{-1}\mathbf{p}-\eta
$$
satisfying the Hamilton-Jacobi equation.

Einstein seems to change his mind in the \emph{Nachtrag zur Korrektur}, p.~91:
\begin{quote}
$\mathbf{LB}$: Liefert eine Bewegung ein $p_i$-Feld, so besitzt dieses notwendig ein Potential $J^*$.
\end{quote}
Whatever he means has to do with \S~4 of \emph{Quantensatz}: ``die zweite der in \S~4 angegebenen Bedingungen für die Anwendbarkeit der Formel 11) stets von selbst erfüllt sein muß [\,\dots]''; at the end of \S~4 we discover that
\begin{quote}
Die Anwendung der Quantenbedingung 11) verlangt, daß derartige Bahnen existieren, daß \so{die einzelne Bahn} ein $p_i$-Feld bestimmt, für welches ein Potential $J^*$ existiert.
\end{quote}
To try to make sense of this we can start with the primitive notion of \emph{Bewegung}: a motion, a trajectory in configuration space---however it may be defined or generated. Returning to the above classification, the motion can be periodic (case [1] or [2]) or not (case [3]). If it is periodic, it will assign to no point $\mathbf{q}$ (or neighbourhood) of $\mathcal{Q}$ more than finitely many momenta $\mathbf{p}_1,\dots,\mathbf{p}_N$. That's what Einstein means by a $p_i$-\emph{Feld}: an assignment of at most finitely many momenta to (certain) points of $\mathcal{Q}$. If the momentum is \emph{finitely} multi-valued here and there, the configuration space can be enlarged to restore single-valuedness---but not if it never closes.

There remains the issue of how a single \emph{Bewegung}, confined as it is to a one-dimensional manifold, can yield a \emph{Feld}, a field on an $l$-dimensional manifold. Since Einstein goes to the trouble of emphasising \so{die einzelne Bahn} with \so{Sperrdruck}, he really does seem to mean a \emph{single} trajectory. He may simply be unaware of the problem; or perhaps it somehow doesn't bother him and he just ignores it; or perhaps he has a way of dealing with it. But if he does, it would surely have to involve a \emph{congruence}, a \so{Schar}, somehow or other---despite the arresting \so{Sperrdruck} of \so{die einzelne Bahn}.

Indeed Einstein's use of Hamilton-Jacobi theory is most peculiar, perhaps even contradictory or downright wrong. It is a theory that undeniably involves \emph{congruences} of trajectories. In \S~3 of \emph{Quantensatz} and in (1917c), Einstein considers the entire congruence; but elsewhere in \emph{Quantensatz} he seems to use the same theory to produce a \emph{single} trajectory. The integral (\ref{integral}) only makes invariant sense if $\mathbf{p}$ is defined \emph{everywhere} on $\mathcal{Q}$, not just on a single trajectory---Einstein accordingly speaks of an entire $p_i$-\emph{Feld}. But Einstein's whole analysis and classification of trajectories (our cases [1]-[3])---which would get confused, perhaps even undermined, by congruences---seems to depend on \emph{single} trajectories. We may simply have a case of Einstein wanting to have his cake and eat it.

Returning to $\mathbf{LB}$, the issue of a Hamilton-Jacobi potential $J^*$ or of a vanishing curl $d\mathbf{p}$ does not even arise with a single trajectory. With a whole congruence of motions radiated from an initial surface, the curl $d\mathbf{p}$ automatically vanishes if the propagation is orthogonal, as in Schrödinger's construction. If the propagation is only transversal, as in \S~3 of \emph{Quantensatz}, the curl has to vanish for the motions to admit a Hamilton-Jacobi potential $J^*$.

\section{Einstein's integrability theorem}\label{Liouville}
In the second-last paragraph of the \emph{Nachtrag zur Korrektur} (1917a, pp.~91-2) Einstein formulates an integrability theorem that deserves attention:
\begin{quote}
$\mathbf{IT}_Q$: Existieren $l$ Integrale der $2l$ Bewegungsgleichungen von der Form
\begin{equation}\label{NachtragTheorem}
R_k(q_i,p_i)=\textrm{konst.,}
\end{equation}
\foreignlanguage{german}{wobei die $R_k$ algebraische Funktionen der $p_i$ sind, so ist $\sum_ip_idq$ immer ein vollständiges Differential, wenn man die $p_i$ vermöge (\ref{NachtragTheorem}) durch die $q_i$ ausgedrückt denkt}.
\end{quote}
Or in Einstein's letter (1917b) to Ehrenfest:
\begin{quote}
$\mathbf{IT}_E$: Es liege ein Problem vor, bei dem soviel Integrale
$$
L\textrm{(}q_{\nu},p_{\nu}\textrm{)}\hspace{1pt}=\hspace{1pt}\textrm{konst}
$$
existieren, als Freiheitsgrade. Dann können die Impulse als (mehrwertige) Funktionen der $q_\nu$ ausgedrückt werden. Andererseits erfülle die Bahnkurve einen gewissen $q_\nu$-Raum vollständig, sodass sie jeden Punkt desselben beliebig nahe kommt. Dann liefert die Bahn des Systems im $q_\nu$-Raum ein Vektorfeld der $p_\nu$.
\end{quote}
Here there are already elements of the theorem now attributed to Liouville\footnote{See \citet{Appell1} pp.~576ff, \citet{Appell} pp.~437ff, \citet{FasanoMarmi} \S11.4.} and Arnol’d.\footnote{See \citet{FasanoMarmi} \S11.5, \citet{Graffi04} \S1.7.1, \citet{Graffi} \S2, \citet{Lowenstein} pp.~56ff.}\color{black}

\subsection{Liouville-Arnol'd}
We can begin anachronistically with (features of) the Liouville-Arnol’d theorem, and then try to understand how much of it is already in \citet{Quantensatz, Ehrenfest}. The theorem uses $l$ functions
$$
F_i:\mathit{\Gamma}\rightarrow\mathbb{R}
$$
on an $2l$-dimensional phase space $\mathit{\Gamma}$ to reduce the number of dynamically relevant dimensions from $2l$ to $l$; in the sense that by $l-1$ intersections of level surfaces of appropriately compatible and independent functions it confines the dynamics to an $l\textrm{-dimensional}$ manifold. A function $F$ foliates the phase space into ($2l-1$)-dimensional level surfaces on which $F=\textrm{const}.$; specifying a value $f$ of  $F$ already eliminates one dimension by fixing a level surface. But the theorem concerns dynamics; the issue is whether a given dynamics\footnote{See \citet{Arnold} \S8.1.3, \citet{SternbergGuillemin} p.~88, \citet{Graffi04} p.~51.}
$$
X_G=(dG)^{\sharp}=\omega^{\sharp}(dG)
$$
is \emph{tangent} to the level surfaces $F=\textrm{const}.$, rather than transversal to them. If the dynamics $X_G$ were transversal to the level surfaces of $F$, the dimension lost by choosing a level surface would be thus restored, with no net progress in the effort to eliminate dimensions.

The relevant notion is Poisson compatibility\footnote{See \citet{Graffi04} p.~50.}
$$
\{F,G\}=0\textrm{,}
$$
which can either be understood as compatibility between the dynamics $X_F$ generated by $F$ and the level surfaces of $G$, or the other way around. $\{F,G\}$ vanishes if the vector field $X_F$ is tangent to the level surfaces of $G$, in other words if the graph of each integral curve of $X_F$ is confined to a level surface of $G$.
We can take the first function $F_1$ to be the Hamiltonian $\mathscr{H}$, whose generic value $E$ singles out a ($2l-1$)-dimensional energy surface. The two compatibilities
$$
\{\mathscr{H},F_2\}=0=\{\mathscr{H},F_3\}
$$
and values $F_2=f_2$, $F_3=f_3$ only eliminate both dimensions  (and not just one) if $dF_2$ and $dF_3$ are independent; if $dF_2$ and $dF_3$ were parallel, $F_2$ and $F_3$ would have the same level surfaces, which would be redundant.\footnote{In the aforementioned `effort to eliminate dimensions,' Poisson incompatibility $\{F,G\}\neq 0$ would be \emph{counter}productive (reversing progress already made by effectively \emph{restoring} an eliminated dimension), whereas linear dependence $dF=kdG$ would be merely \emph{un}productive (producing neither loss nor gain, leaving the number of dimensions unchanged).} \emph{Complete integrability} is given by $l$ values
$$
f=(f_1,\dots,f_l)
$$
of the compatible, independent functions $\mathfrak{F}_l=\{F_1,\dots,F_l\}$, which eliminate $l$ of the initial $2l$ dimensions of phase space, leaving an $l$-dimensional manifold $\mathcal{M}_f$. 

So far we have little more than a number $l$ of dimensions. Without compactness, the $l$-dimensional manifold $\mathcal{M}_f$ could be a product of lines and loops; compactness rules out the lines, leaving a torus $\mathfrak{T}^l$, a product of $l$ topological circles. It is worth noting that Arnol'd and Einstein obtain their tori in different ways: Arnol'd by imposing compactness, Einstein by \emph{Riemannisierung}, to eliminate self-intersections. The two ways, however different, are not unrelated: if the $l$-dimensional manifold $\mathcal{M}_f$ were a product of $l$ lines, it would amount to $\mathbb{R}^l$; compactness prevents immersion in $\mathbb{R}^l$ by `swelling' $\mathcal{M}_f$. A two-dimensional torus $\mathfrak{T}^2$, for instance, is an `enlarged' two-dimensional configuration space inasmuch as it cannot be embedded in the plane. Suppose for definiteness that $\mathfrak{T}^2$ is not just a `topological torus' (a product of two loops) but a `rigid torus' (a product of two rigid circles $\mathbb{S}^1$)---literally a doughnut, with a shape and a size, embedded in $\mathbb{R}^3$ parallel to the $xy$ plane. Take the simplest possible motion, given by fixed rates of rotation around both circles of such a $\mathfrak{T}^2$: projected onto the $xy$ plane it would intersect itself (at regular intervals). That's how compactness is related, albeit indirectly, to Einstein's \emph{Riemannisierung}.

The distinction between rigid and topological tori, introduced above for mere definiteness of representation, can actually prove quite relevant. According to the Liouville-Arnol'd theorem, a completely integrable (and entirely compact) dynamics can be represented as $l$ constant rates
\begin{equation}\label{rates}
\dot{\varphi}_i=\omega_i=\frac{\partial \mathscr{H}}{\partial J^*_i}
\end{equation}
of rotation on an $l$-dimensional torus $\mathfrak{T}^l$, where the $l$ angles\footnote{See \citet{SternbergGuillemin} pp.~356ff, \citet{Lowenstein} pp.~68ff.} $\varphi_i$ with linear evolutions
$$
\varphi_i\textrm{(}t\textrm{)}=\omega_it+\varphi_i\textrm{(}0\textrm{)}
$$
are canonically conjugate
\begin{equation}\label{conjugacy}
\{\varphi_i,J^*_j\}=\delta_{ij}
\end{equation}
to the actions $J^*_i$; $i,j=1,\dots,l$.

Einstein seems to construct a \emph{topological}, $l$-dimensional torus (topologically equivalent to a rigid torus) to resolve self-intersections; his $l$ integrals (\ref{integral}) are undeniably \emph{action} integrals; but that's not enough to provide the `symplectic' rigidity that turns a topological torus into a rigid one. To do so, Einstein would have needed something along the lines of (\ref{rates}) or (\ref{conjugacy})---which were by no means obvious in 1917, especially to a physicist, and cannot be taken for granted. To obtain a rigid torus one has to recognise that the angles $\varphi_i$ yielding the constant frequencies $\omega_i$ are canonically conjugate to the actions $J^*_i$; but nowhere does Einstein betray such symplectic awareness. The rigid torus of the Liouville-Arnol'd theorem is really quite different from Einstein's merely topological torus: it has a metrically definite shape and is twice as big, with $2l$ degrees freedom ($l$ areas and $l$ angles), not just $l$ ($l$ arbitrary parameters along $l$ loops). Even if constant motions on a rigid torus are ultimately needed to make sense of his intuitive analysis of integrability (closed \emph{vs}.\ space-filling motions), that distinction alone hardly warrants the attribution of so much definite structure to Einstein's loose, highly topological constructions.

But let us nonetheless consider a rigid torus, which with little loss of generality can be taken to be two-dimensional, with frequencies $\omega_1$, $\omega_2$. The motion can in any case be confined to a one-dimensional manifold; the issue is its length (finite or not), its topology (closed or open)---whether we have a \emph{Bewegung in exakt geschlossener Bahn}, whose \emph{Punkte bilden ein Kontinuum von nur einer Dimension}. The relevant criterion is rational dependence: if ${\omega_1}/{\omega_2}$ is rational, the motion can be confined to a one-dimensional \emph{Kontinuum} which, being \emph{geschlossen}, is of finite length; whereas if $\omega_1/\omega_2$ is irrational the motion can still be confined to a one-dimensional \emph{Kontinuum}, but not of finite length.\footnote{See \citet{Born33} pp.~80ff, \citet{FasanoMarmi} \S11.7.} Einstein clearly understands the geometrical significance of confining motion to intersections of level surfaces of appropriately compatible and independent functions; what he doesn't mention is the \emph{numerical}---rational \emph{vs}.\ irrational---rather than set-theoretical (or `manifold-theoretical') character of the last step, needed to bring the (appropriately finite) dimensions down to \emph{one}.

\subsection{Integrability in \emph{configuration} space}
To understand Einstein's integrability theorem(s) $\mathbf{IT}$ ($\mathbf{IT}_Q$ \& $\mathbf{IT}_E$) one has to bear in mind above all that he's in \emph{configuration} space $\mathcal{Q}$ and not in phase space $\mathit{\Gamma}$. Again, throughout (1917a,b,c), Einstein is clearly and explicitly\footnote{See footnote \ref{ConfigurationFootnote}.} in the $l$-dimensional space he calls $q$-\emph{Raum}; this is especially evident at the bottom of p.~87 (1917a):
\begin{quote}
\foreignlanguage{german}{Die Bahnkurve läßt sich ganz in einem Kontinuum von weniger als $l$ Dimensionen unterbringen.}
\end{quote}
The kind of integrability he has in mind, taken as far as possible, would confine the \emph{Bahnkurve} to a \emph{one}-dimensional submanifold of $\mathcal{Q}$---not of $\mathit{\Gamma}$:
\begin{quote}
\foreignlanguage{german}{Hierzu gehört als spezieller Fall derjenige der Bewegung in exakt geschlossener Bahn. [\,\dots] die Bahn ist dann eine geschlossene, ihre Punkte bilden ein Kontinuum von nur einer Dimension.}\footnote{\emph{ibid}.\ pp.~87-8}
\end{quote}

Once we've fixed $\mathbf{q}\in\mathcal{Q}$, the functions $\mathfrak{F}_l=\{F_1,\dots,F_l\}$ depend on momentum alone. If the function $F_k$ is, say, quadratic in $\mathbf{p}$, a value $f_k$ determines an ellipsoid
$$
\Sigma(f_k)\subset\mathbb{T}_\mathbf{q}\mathcal{Q}
$$
at $\mathbf{q}$ (and otherwise a more general surface). To simplify, ruling out intractable pathologies, Einstein requires $\mathfrak{F}_l$ to be algebraic (not necessarily quadratic) functions of mo\-mentum---he needs a class of functions \emph{that's small enough to allow his procedure to work}. The $l$ values $f_1,\dots,f_l$ of the functions $\mathfrak{F}_l$ of $\mathbf{p}$ at $\mathbf{q}\in\mathcal{Q}$ would confine $\mathbf{p}$ to the intersection
$$
\Sigma(f_1)\cap \cdots \cap \Sigma(f_l)\subset\mathbb{T}_\mathbf{q}\mathcal{Q}.
$$
The point of the theorem is not to express, given a set $\mathfrak{F}_m$, the exact number $\#\mathbf{p}\textrm{(}\mathbf{q}\textrm{)}$ of momenta
$$
\mathbf{p}\in\Sigma(f_1)\cap \cdots \cap \Sigma(f_l)
$$
at $\mathbf{q}\in\mathcal{Q}$ as a function of $m$. Einstein only wants to know whether a certain set $\mathfrak{F}_m$, a certain integer $m$, corresponds to \emph{finitely} many momenta at $\mathbf{q}$ or \emph{infinitely} many momenta. And concerning $m$ itself, he's mainly (perhaps even \emph{only}) interested in whether $m<l$ or $m=l$. Even if that simplifies matters considerably, the logic of $\mathbf{IT}$ remains somewhat ambiguous. The clear part of the logic is that an \emph{incomplete} set $\mathfrak{F}_m$ ($m<l$) allows \emph{infinitely} many momenta $\mathbf{p}$ at the \emph{Stelle}---the incompleteness rules out \emph{finitely} many momenta (and hence a closed, periodic motion). A \emph{complete} set $\mathfrak{F}_l$ could in principle still allow finitely many \emph{or infinitely many} momenta. Without being entirely explicit, Einstein pretty clearly suggests that a complete set $\mathfrak{F}_l$ rules out infinitely many momenta. I read him as saying:
$$
\textrm{\emph{complete} set } \mathfrak{F}_l \Rightarrow \textrm{\emph{periodic} motion;}
$$
\emph{motion is closed, periodic, `totally' integrable (as opposed to `space-filling') if and only if the set of integrals is complete}.

Returning to $\mathbf{IT}_Q$, Einstein mentions that the momentum (\ref{momentum}) would be a \emph{vollständiges Differential}. I think the point isn't \emph{exact \emph{vs}.\ $\kappa$losed}, but rather that once non-periodic (space-filling) motion is ruled out, with its infinitely many momenta, the momentum just \emph{makes sense}, having only finitely-many values---or \emph{just one} at every point of an appropriately enlarged configuration space.

Since $\mathbf{IT}_E$ is somewhat choppy, its four sentences [S1]-[S4] can be looked at one by one. In my notation:
\begin{enumerate}[label={[S\arabic*]}]
\item The set $\mathfrak{F}_m$ at $\mathbf{q}\in\mathcal{Q}$ is assumed complete: $m=l$.
\item Since $\mathfrak{F}_m$ is complete, there are at most \emph{finitely} many momenta $\mathbf{p}$ at $\mathbf{q}$.
\item Were $\mathfrak{F}_m$ \emph{in}complete ($m<l$), the motion would fill space.
\item Since $\mathfrak{F}_m$ \emph{is} complete, the (periodic) motion yields a momentum (co)vector field (with \emph{finitely} many values) on $\mathcal{Q}$.
\end{enumerate}
Einstein really does seem to say that a complete set $\mathfrak{F}_l$ guarantees a closed, periodic motion.\color{black}

\section{Final remarks}
Einstein situates his mechanics in the $q_i$-\emph{Raum} he so often refers to; symplectic abstractions are foreign to him.
The `complete' integrability one comes across in the modern literature is hardly complete by Einstein's standards, being compatible with both periodic and space-filling motions. The more extreme kind of integrability Einstein has in mind corresponds to periodic, not space-filling motions: \emph{tertium non datur}. He needs a self-intersecting motion assigning more than one momentum here and there---to justify the enlarged configuration space whose topological peculiarities are captured by the homotopy classes he integrates over---but not the \emph{infinitely many} momenta of space-filling motion; his \emph{Riemannisierung} is necessarily finite. And once on the torus, where the greater invariance of Einstein's integrals $\langle\mathbf{p},\mathbb{H}_i\rangle$ is evident, the score against Sommerfeld is (an admittedly mathematical, nonempirical) 2-1.
\vspace{10pt}

\noindent I thank Ermenegildo Caccese, Sandro Graffi, Stefano Marmi and Nic Teh for many useful conversations; and audiences in Urbino, at Paris Diderot and Notre Dame for valuable feedback.

\end{document}